\documentclass[letterpaper]{JHEP3}
\usepackage{epsfig,psfrag}
\usepackage{amsmath,amssymb,exscale}
\usepackage{amscd,shadow,fancybox,axodraw,booktabs}

\newcommand{\be}{\begin{equation}}
\newcommand{\ee}{\end{equation}}
\newcommand{\bear}{\begin{eqnarray}}
\newcommand{\eear}{\end{eqnarray}}
\newcommand{\ba}{\begin{array}}
\newcommand{\ea}{\end{array}}
\newcommand{\lae}{\begin{array}{c}\,\sim\vspace{-21pt}\\<
\end{array}}

\newcommand{\tr}{{\rm{Tr}}}

\title{Renormalization of the $S$ Parameter in Holographic Models of Electroweak 
Symmetry Breaking
\\ { $\; $ } \\ }

\author{Gustavo Burdman and  Leandro Da Rold\\
Departamento de F\'{i}sica Matem\'{a}tica\\ 
Instituto de F\'{i}sica, Universidade de S\~{a}o Paulo,
\\ R. do Mat\~{a}o 187, S\~{a}o Paulo, SP 05508-900, Brazil \\ 
\\ \\
\email{burdman@if.usp.br, daroldl@fma.if.usp.br} \\ }

\abstract{
We show that the $S$ parameter is not finite in theories of 
electroweak symmetry breaking in a slice of anti--de Sitter 
five-dimensional space, with the light fermions localized in the ultraviolet. 
We compute the one-loop contributions to $S$ from the Higgs sector and show that  
they are logarithmically dependent on the cutoff of the theory. 
We discuss the renormalization of $S$, as well as 
the implications for bounds from electroweak precision 
measurements on these models. We argue that, although in principle the 
choice of renormalization condition could eliminate the $S$ parameter constraint, 
a more consistent condition would still result in a large and positive
$S$. On the other hand, we show that the dependence on the Higgs mass in $S$ 
can be entirely  eliminated by the renormalization
procedure, making it impossible in these theories to extract a Higgs
mass bound from electroweak precision constraints. 
}

\preprint{hep-ph/yymmnnn}


\begin{document}


\section{Introduction} \setcounter{equation}{0}
\label{intro}

Although the standard model (SM) is an extremely successful description of the
electroweak interactions, the instability of the weak scale under radiative 
corrections leads us to believe that there should be physics beyond the SM at an energy 
scale not far beyond the TeV. The origin of electroweak symmetry breaking (EWSB) 
as well as of fermion masses, might be associated with this new dynamics. 
A proposal for stabilizing the weak scale using a theory with one compact extra dimension 
with a non-factorizable, Anti de Sitter metric~\cite{rs}, the Randall-Sundrum (RS) model, 
can be 
thought of as dual to a strongly coupled four-dimensional theory with
a large number of colors~\cite{holorefs}. 
The slice of AdS$_5$ is 
defined by an ultra-violet (UV) fixed point located at the Planck scale, 
$M_P$, and an infra-red (IR) one, with an exponentially suppressed scale which is identified as 
the TeV scale. The 5D metric in conformal coordinates is given by:
\begin{equation}
ds^2=\left(\frac{1}{kz}\right)^2(\eta_{\mu\nu}dx^\mu dx^\nu-dz^2)
\end{equation}
where $k$ is the AdS$_5$ curvature. 
 This spacetime has two 4D boundaries at 
$z_0=1/k\sim 1/M_{Pl}$ and $z_1\sim 1/$TeV, respectively 
the UV and IR boundaries.

In order to stabilize the weak scale, the Higgs field must be localized at or near the 
TeV brane. This is not the case with the rest of the fields, which can then propagate in the 
AdS$_5$ bulk. The theories built this way, bulk AdS$_5$ models, not only avoid potentially
troublesome higher dimensional operators suppressed only by the TeV scale, but also allow 
for a natural explanation of the fermion mass hierarchy~\cite{gp,gn,cn}. 

There are several possibilities for building models of electroweak symmetry breaking in 
AdS$_5$. The basic elements for building a successful theory include 
the choice of the bulk gauge group, the zero-mode fermion localization and the 
dynamical mechanism for localizing the Higgs field on or near the TeV brane. 
The bulk gauge symmetry must be enlarged with respect to the SM in order to include 
isospin symmetry and avoid tree level contributions to the $T$ parameter. 
A minimal extension~\cite{adms} 
is $SU(2)_L\times SU(2)_R\times U(1)_X$, broken by boundary conditions either 
to the SM gauge group $SU(2)_L\times U(1)_Y$ or directly to $U(1)_{\rm EM}$ as in 
Higgsless models~\cite{higgsless}. 
In order to naturally address the fermion mass hierarchy, light fermions must be localized 
close to the UV boundary, and heavier fermions, such as the top quark, must be localized 
towards the IR brane to have a significant Yukawa coupling to the Higgs field~\cite{gp,gn}. 
Finally, the IR localization of the Higgs can be dynamically achieved in specific models 
of EWSB. For instance, in a Gauge-Higgs unification model~\cite{cnp,acp} in AdS$_5$, the 
Higgs arises from the $A_5$ components of a gauge field and is naturally localized
towards the TeV brane, as required to solve the hierarchy problem; whereas the inclusion of 
a fourth-generation highly localized towards the IR brane can result in a condensation of some of the 
fourth-generation zero modes and therefore in a Higgs localized near the IR~\cite{fgen}.  
IR localization of the Higgs can even be achieved in soft-wall models
without an IR brane, as it is shown in Reference~\cite{afmpv} in a 
Gauge-Higgs unification model. 

These bulk AdS$_5$ models of EWSB and fermion masses can be thought of as duals 
of some strongly coupled 4D theory. They all share a common problem regarding 
electroweak precision constraints: a tree-level $S$ parameter. This is
approximately given by~\cite{adms,sispositive}
\be
S_{\rm tree} \simeq 2\pi \,v^2\,z_1^2~,
\ee
where $v\simeq 246~$GeV is the vacuum expectation value of the Higgs
field and we took the limit $vz_1\ll 1$\footnote{In Gauge-Higgs unification models 
there is typically an additional suppression given by $(v/f_\pi)^2$, the ratio of the Higgs VEV 
to the symmetry breaking scale~\cite{acp}.}. 
For a TeV scale IR brane this results in $S_{\rm tree} \simeq
0.3$, in contradiction with current electroweak constraints~\cite{pdg}. 
It is possible to avoid this problem by de-localizing fermions 
~\cite{adms,delocal}. But in doing so, we would loose one of the most
interesting features of these theories, namely a natural way of
generating the fermion mass hierarchy. In this paper, we will restrict
ourselves to models with light fermions localized near the UV boundary. 

The presence of the tree-level $S$ parameter in all bulk AdS$_5$
models poses a very stringent constraint on them. It suggests
that it would be of interest to study the loop contributions to it. 
In this paper we compute the one loop contributions to the $S$
parameter in these models coming from loops involving the Higgs sector.
We will show that the one loop contributions to $S$ are
logarithmically divergent and therefore require that $S$ be properly
renormalized. We argue that similar divergences are expected in the fermion and gauge
boson loops. 
The fact that $S$ is logarithmically sensitive to the cutoff should
not be completely surprising. From the point of view of the  5D theory,
this is the cutoff   of the non-renormalizable theory,  properly warped down. 
On the other hand, in the 4D holographic picture, this cutoff corresponds to the matching 
of the low energy effective theory and the 4D strongly coupled CFT. 
A more subtle
question is the choice of a renormalization condition for
$S$. Although the logarithmic divergence is sub-dominant in the large
$N$ expansion, it could be numerically sizable. 
Furthermore, the 
renormalization procedure introduces a scale dependence in the $S$
parameter. 
All in all, the use of the $S$ parameter as a tight constraint on the
mass scale of the Kaluza-Klein (KK) excitations as well as on the 
Higgs mass must be reassessed. 

The plan for the rest of the paper is as follows: in the next Section  
we present the setup of AdS$_5$ bulk models and derive the low energy
effective theory obtained after integrating out the 5D bulk; in
Section~\ref{higgsconts} we compute the one-loop contributions of the 
Higgs sector to the $S$ parameter in the effective theory and discuss
the renormalization procedure. Finally, in Section~\ref{conclusions} 
we discuss our results and conclude.

\section{Electroweak Symmetry Breaking in AdS$_5$}
\label{ewsb5d}

We consider a 5D model in a slice of AdS, with the gauge symmetry
SU(2)$_L\times$SU(2)$_R\times$U(1)$_X$, broken to the SM in the UV
boundary. The 5D action is given by:
\begin{equation}
S=\int d^4x\int dz\,\sqrt{g}\, \left[-\frac{1}{4} L^a_{MN}L^{aMN}
-\frac{1}{4}R^a_{MN}R^{aMN}-\frac{1}{4}X_{MN}X^{MN}\right]\, ,
\label{l5}
\end{equation}
where
$L_{MN}^a$, $R_{MN}^a$ and $X_{MN}$ are the $SU(2)_L$, $SU(2)_R$ and
$U(1)_X$ field strengths,  and $g$ is the determinant of the metric.

The Higgs field transforms as $(\bf{2},\bf{2})_{\bf{0}}$ under the
gauge symmetry, 
\begin{equation}
H=
\frac{1}{\sqrt{2}}
\left(\begin{array}{ccc}
v+h+i\phi_3 &~ ~ &i(\phi_1-i\phi_2)\\ i(\phi_1+i\phi_2) &~ ~ & v+h-i\phi_3
\end{array}\right) \, 
\end{equation}
 and it is localized near the IR brane by some suitable
mechanism, such as in Gauge-Higgs unification~\cite{acp} or the condensation of a
fourth-generation zero-mode fermion~\cite{fgen}. Here it suffices to
assume an effective localization on the IR boundary as given by  
\be
S_{\rm IR} = \int d^4x\int dz\,\delta(z-z_1)\,
\sqrt{g_{\rm IR}}\left[\frac{1}{2}\tr|D_\mu H|^2-V(H)\right]\, ,\label{sir}
\ee
with $g_{\rm IR}$  the induced metric in the IR boundary and $V(H)$
the usual renormalizable Higgs potential.
The covariant derivative acting on the scalar field is defined as:
\begin{equation}
D_\mu H= \partial_\mu H-ig_5L_\mu H+i\tilde{g}_5H R_\mu\, , 
\end{equation}
where $g_5$ and $\tilde{g}_5$ are the $SU(2)_L$ and $SU(2)_R$ 5D gauge
couplings, respectively. 
As usual, in order to obtain a canonically normalized Higgs kinetic
term, we rescale the Higgs field by $H\to (1/kz_1)\,H$.

\subsection{The Low Energy Effective Theory}
\label{leet}
The presence of the 5D bulk affects the couplings of gauge bosons to
the Higgs sector, as well as to fermions. 
In order to compute the one loop contributions to electroweak
precision constraints, we will integrate out the 5D bulk and obtain
a low energy theory containing the zero-mode gauge bosons and the
Higgs. We will use the holographic approach to obtain the resulting
low energy effective theory, 
separating the UV degrees of
freedom. This is useful  
since the UV boundary and the bulk respect different symmetries.

The Higgs VEV $\langle H\rangle=v {\bf I}/\sqrt{2}$ breaks the
SU(2)$_L\times$SU(2)$_R$ symmetry down to SU(2)$_V$. Therefore it is convenient
to work in the vector and axial-vector basis in the bulk, defined by
\begin{eqnarray}
V_M&=&\frac{1}{\sqrt{g_{5}^2+\tilde{g}_5^2}}\big(\tilde{g}_{5}L_M+\tilde{g}_5R_M)\, ,\nonumber\\
A_M&=&\frac{1}{\sqrt{g_{5}^2+\tilde{g}_{5}^2}}\big(g_{5}L_M-\tilde{g}_{5}R_M)\, .
\end{eqnarray}
We add the gauge fixing term:
\begin{equation}
\begin{aligned}
{\cal L}_{GF}^V&=-\frac{1}{kz\,\xi_V}\tr\left[\partial_{\mu}V_{\mu}-z\,\xi_V\partial_5
(V_5/z)\right]^2\, ,
\end{aligned}
\end{equation}
where $\partial_5$ is the derivative with respect to the $z$
coordinate, and there will be similar terms for $A_M$ and $X_M$. We will take the limit
$\xi_{V,A,X}\rightarrow\infty$, and obtain
$\partial_5(V_5/(kz))=\partial_5(A_5/(kz))=\partial_5(X_5/(kz))=0$.
After integration by parts the quadratic term for $V_\mu$ in the
5D Lagrangian is
\begin{equation}\label{L2}
{\cal L}=\frac{1}{kz}
\tr\left\{V_{\mu}\left[(\partial^2-z\partial_5 (1/z)\,\partial_5) \eta_{\mu\nu} -
\partial_{\mu}\partial_{\nu}\right]V_{\nu}\right\} + \dots \, ,
\end{equation}
and similarly for $A_\mu$ and $X_\mu$.

Also left from the integration by parts are the following boundary
terms: 
\begin{equation}
{\cal L}_{\rm{bound}}=\frac{1}{kz}
\tr\left[V_{\mu}\partial_5V_{\mu}-2V_{\mu}\partial_{\mu}V_5
+ A_{\mu}\partial_5A_{\mu}-2A_{\mu}\partial_{\mu}A_5
+ X_{\mu}\partial_5X_{\mu}-2X_{\mu}\partial_{\mu}X_5\right]\Big|^{z_1}_{z_0} \, .
\end{equation}
Since the IR-localized Higgs acquires a VEV, its 
kinetic term mixes $A_\mu$ with the Nambu-Goldstone bosons (NGBs) $\phi_i$
(i=1,2,3). 
We then add an additional  gauge fixing term on the IR boundary
\begin{equation}
\begin{aligned}
{\cal L}_{GF,{\rm IR}}^A&=-\frac{1}{\xi_{A,{\rm IR}}}
\tr\left(\partial_{\mu}A_{\mu}
-\frac{\xi_{A,{\rm
IR}}}{2}\sqrt{(g_{5}^2+\tilde{g}_{5}^2)k}\,v\sigma^i
\phi_i\right)^2\Big|_{z_1}
\, .
\end{aligned}
\end{equation}
We choose $\xi_{A,{\rm IR}}=0$.  
We then solve the bulk equations of motion obtained from (\ref{L2}), with 
the following boundary conditions on the IR:
\begin{eqnarray}
& &\partial_5 V_\mu|_{z_1}=V_5|_{z_1}=\partial_5 X_\mu|_{z_1}=X_5|_{z_1}=0 \, ,\\
& &\left(\frac{1}{kz}\partial_5+\frac{g_{5}^2+\tilde{g}_5^2}{4}v^2\right)A_\mu|_{z_1}
=A_5|_{z_1}=0 \, .
\end{eqnarray}
The solutions can be written as 
\begin{eqnarray}
V_\mu(p,z)=\sqrt{k}\,V_\mu^0(p)\,f_V(p,z) \, ,
& & A_\mu(p,z)=\sqrt{k}\,A_\mu^0(p)\,f_A(p,z) \nonumber\\
 & &~\nonumber\\
X_\mu(p,z)&=&\sqrt{k}\,X_\mu^0(p)\,f_V(p,z) \, ,
\end{eqnarray}
\noindent with $f_{V,A}$ defined by:
\begin{eqnarray}
f_V(p,z)&=&\frac{z(J_1(pz)Y_0(pz_1)-Y_1(p z)J_0(pz_1))}
{z_0(J_1(pz_0)Y_0(p z_1)-Y_1(pz_0)J_0(pz_1))}\, ,\\
f_A(p,z)&=&\frac{z[J_1(pz)(pY_0(pz_1)+m_1Y_1(pz_1))-Y_1(pz)(pJ_0(pz_1)+m_1J_1(pz_1))]}
{z_0[J_1(pz_0)(pY_0(pz_1)+m_1Y_1(pz_1))-Y_1(pz_0)(pJ_0(pz_1)+m_1J_1(pz_1))]}~,
\end{eqnarray}
where 
\be
m_1=(g_5^2+\tilde{g}_5^2)k v^2z_1/4~, 
\label{m1def}
\ee
and where 
$V_\mu^0(p)=V_\mu(p,z_0)/\sqrt{k}$ is the UV-boundary value of the
$V_\mu$ field, with analogous definitions for the UV fields of $A_\mu$
and $X_\mu$. In what follows we will drop the
index $0$, and will refer to the UV fields simply as $V_\mu$, $A_\mu$ and $X_\mu$.

The low energy effective theory can then be written in terms of the UV
fields and the IR-localized Higgs. It is obtained by
substituting  the solutions for the UV fields 
back into the action. 
The resulting low energy effective theory comes from the UV
boundary terms, and describes the interactions of the ``elementary'' fields coupled to the 
IR-localized Higgs. These interactions encode the effects of the bulk that was integrated   
out. In order to have the SM gauge field content at low energies, 
we choose the dynamical fields at low energy to be 
the SU(2)$_L\times$U(1)$_Y$ gauge fields
\begin{equation}
L^a_\mu \, , \, a=1,2,3\, ; \qquad B_\mu=\frac{g_{5X}R^3_\mu+\tilde{g}_{5}X_\mu}
{\sqrt{\tilde{g}_{5}^2+g_{5X}^2}} \, ,
\end{equation}
\noindent 
with $g_{5X}$ the 5D $U(1)_X$ gauge coupling,  
whereas the other gauge fields in the UV 
\begin{equation}
R^a_\mu \, , \, a=1,2 \, ; \qquad S_\mu=\frac{\tilde{g}_{5}R^3_\mu-g_{5X}X_\mu}
{\sqrt{\tilde{g}_{5}^2+g_{5X}^2}} \, ,
\end{equation}
are given Dirilichet boundary conditions and are not present in the
effective theory. 
We define the 5D hypercharge coupling constant by:
\begin{equation}
g_{5Y}=\frac{\tilde{g}_{5}g_{5X}}
{\sqrt{\tilde{g}_{5}^2+g_{5X}^2}} \, .
\end{equation}
After integrating out the bulk gauge fields, the momentum-space
quadratic terms in the effective Lagrangian are
\begin{eqnarray}\label{l2effLB}
{\cal L}_{\rm{eff}}^2&=&
\frac{P_{\mu\nu}}{2}\left[L^a_\mu\Pi_L(p^2)L^a_\nu+2L^3_\mu\Pi_{3B}(p^2)B_\nu+B_\mu\Pi_B(p^2)B_\nu\right]
\nonumber\\
&-& \frac{1}{2}h(p^2+m_h^2)h-\frac{1}{2}\phi_i p^2 \phi_i \, ,
\end{eqnarray}
\noindent where the correlators $\Pi_L,\Pi_{3B}$ and $\Pi_B$ are given
by
\begin{eqnarray}
\Pi_L(p^2)&=&\frac{\tilde{g}_5^2\Pi_V+g_5^2\Pi_A}{g_5^2+\tilde{g}_5^2} \,
,\\
\Pi_{3B}(p^2)&=&\frac{g_5\tilde{g}_5g_{5X}}{(g_5^2+\tilde{g}_5^2)\sqrt{\tilde{g}_5^2+g_{5X}^2}}(\Pi_V-\Pi_A)
\label{pi3b}\, ,\\
\Pi_B(p^2)&=&\frac{(g_5^2g_{5X}^2+g_5^2\tilde{g}_5^2+\tilde{g}_5^4)\Pi_V+\tilde{g}_5^2g_{5X}^2\Pi_A}
{(g_5^2+\tilde{g}_5^2)(\tilde{g}_5^2+g_{5X}^2)} \,
,\\
\end{eqnarray}
and $\Pi_{V,A}(p^2)$ are the vector and axial correlators, defined as
\begin{eqnarray}
\Pi_V(p^2)&=&-\frac{p[J_0(pz0)Y_0(pz_1)-Y_0(pz_0)J_0(pz_1)]}
{z_0[J_1(pz_0)Y_0(p z_1)-J_0(pz_1)Y_1(pz_0)]}\, ,\\
\Pi_A(p^2)&=&-\frac{pJ_0(pz_0)[pY_0(pz_1)+m_1Y_1(pz_1)]-pY_0(pz_0)[pJ_0(pz_1)+m_1J_1(pz_1)]}
{z_0J_1(pz_0)[pY_0(pz_1)+m_1Y_1(pz_1)]-z_0Y_1(pz_0)[pJ_0(pz_1)+m_1J_1(pz_1)]}\,.
\label{avcorrel}
\end{eqnarray}
The tree-level contribution to $S=-16\pi/(g g')\Pi_{3B}'(0)$, 
can already be obtained from the momentum-dependent correlator in (\ref{pi3b}). 
Defining the 4D gauge couplings by 
\be
g_5^2\simeq\frac{1}{k}\,\log\frac{z_1}{z_0}\,g^2 , \qquad
g_{5Y}^2 \simeq\frac{1}{k}\,\log\frac{z_1}{z_0}\,g'^2~,
\ee
where have discarded terms of order ${\cal O}(vz_1)^2$, one obtains
\begin{equation} 
S_{\rm{tree}}=4\pi v^2z_1^2\,\frac{32+3(g_5^2+g_{5Y}^2)v^2z_1^2}
{(8+(g_5^2+g_{5Y}^2)v^2z_1^2)^2}
\simeq 2\pi v^2z_1^2 \, ,
\label{stree}
\end{equation}
where we have taken the limit $vz_1\ll 1$ to obtain the last expression.

In the absence of new terms localized on the UV boundary, the
propagators of the UV fields are given by the inverse of the
correlators. 
In  the diagonal basis $\{\gamma_\mu,Z_\mu\}$ we have 
\begin{equation}
\gamma_\mu=\frac{\tilde{g}_5g_{5X}L^3_\mu+g_5\sqrt{\tilde{g}_5^2+g_{5X}^2}B_\mu}
{[g_5^2(\tilde{g}_5^2+g_{5X}^2)+\tilde{g}_5^2g_{5X}^2]^{1/2}} \, ,
\qquad 
Z_\mu=\frac{g_5\sqrt{\tilde{g}_5^2+g_{5X}^2}L^3_\mu-\tilde{g}_5g_{5X}B_\mu}
{[g_5^2(\tilde{g}_5^2+g_{5X}^2)+\tilde{g}_5^2g_{5X}^2]^{1/2}} \, ,
\end{equation}
with correlators given by 
\begin{equation}\label{fotonzetacorrel}
\Pi_\gamma=\Pi_V \, ,
\qquad 
\Pi_Z=
\frac{\tilde{g}_5^4\Pi_V+(\tilde{g}_5^2g_{5X}^2+g_5^2\tilde{g}_5^2+g_5^2g_{5X}^2)\Pi_A}
{(g_5^2+\tilde{g}_5^2)(\tilde{g}_5^2+g_{5X}^2)} \, .
\end{equation}
Finally, the spectrum of vector resonances, corresponding to the KK
spectrum,  is given by the zeroes of  
$\Pi_\gamma(p^2)$, $\Pi_Z(p^2)$ and $\Pi_L(p^2)$.

\subsection{Gauge-Higgs interactions}
\label{ghint}
In order to compute the one-loop corrections to the $S$ parameter
coming from the Higgs sector, we need the interactions of the
gauge bosons and the Higgs in the low energy effective theory. 
The interactions of interest are the cubic interactions described by 
\begin{eqnarray}\label{Higgscouplings1}
{\cal L}^3_{\rm{eff}}&=&
\frac{g}{2} L^{1\,\mu}(p) 
[c_A(p)\,h\overleftrightarrow{\partial_\mu} \phi_1+
c_V(p)\,\phi_3\overleftrightarrow{\partial_\mu} \phi_2]\nonumber\\
&+&
\frac{g}{2} L^{3\,\mu}(p) 
[c_A(p)\,h\overleftrightarrow{\partial_\mu} \phi_3+
c_V(p)\,\phi_2\overleftrightarrow{\partial_\mu} \phi_1]\nonumber\\
&+&
\frac{g'}{2} B^\mu(p)
[-c_A(p)\,h\overleftrightarrow{\partial_\mu} \phi_3+
\tilde{c}_V(p)\,\phi_2\overleftrightarrow{\partial_\mu} \phi_1]\, , 
\end{eqnarray}
where $c_A(p)$, $c_V(p)$ and $\tilde{c}_V(p)$ have a
non-trivial dependence with momentum and are defined by
\begin{eqnarray}\label{defgammabeta}
c_V(p)=\frac{2\tilde{g}_5^2f_V(p,z)+(g_5^2-\tilde{g}_5^2)f_A(p,z)}{(g_5^2+\tilde{g}_5^2)}\Big|_{z_1}\;,
\;& & \; 
\tilde{c}_V(p)=\frac{2g_5^2f_V(p,z)+(\tilde{g}_5^2-g_5^2)f_A(p,z)}{(g_5^2+\tilde{g}_5^2)}\Big|_{z_1}\;,
\nonumber\\
~  ~ \nonumber\\
c_A(p)&=&f_A(p,z)\Big|_{z_1}\, .
\end{eqnarray}
Taking the limit of $z_1\rightarrow z_0$ we recover the SM 
couplings, with $c_V=\tilde{c}_V=c_A=1$,
but for finite $(z_1-z_0)\sim 1/$TeV the gauge-Higgs couplings are modified with
respect to their SM values. In particular, the fact that $c_V(p)\tilde{c}_V(p)\not =
c_A^2(p)$ in (\ref{defgammabeta}) will result in divergences in
the one loop calculation of the $S$ parameter.

We will also need the quartic interactions given by
\begin{eqnarray}
{\cal L}^4_{\rm{eff}}&=&
\frac{g^2}{8}L^1_\mu(p) L_1^\mu(k)
\Big\{[(2v+h)h+\phi_1^2]c_A(p)c_A(k)+(\phi_2^2+\phi_3^2)c_V(p)c_V(k)\,\Big\} \nonumber\\
&+&
\frac{g^2}{8}L^3_\mu(p) L_3^\mu(k)
\Big\{[(2v+h)h+\phi_3^2]c_A(p)c_A(k)+(\phi_1^2+\phi_2^2)c_V(p)c_V(k)\Big\} \nonumber\\
&+&
\frac{g'^2}{8}B_\mu(p) B^\mu(k)
\Big\{[(2v+h)h+\phi_3^2]c_A(p)c_A(k)+(\phi_1^2+\phi_2^2)\tilde{c}_V(p)\tilde{c}_V(k)\Big\} \nonumber\\
&+&
\frac{gg'}{4} B_\mu(p) L_1^\mu(k) (v+h) \phi_2 \left(\frac{g_5^2c_A(k)}{g_5^2+\tilde{g}_5^2}
[c_V(p)+\tilde{c}_V(p)]+\frac{\tilde{g}_5^2c_A(p)}{g_5^2+\tilde{g}_5^2}[c_V(k)+\tilde{c}_V(k)]\right)
\nonumber\\
&+&
\frac{gg'}{4} B_\mu(p) L_3^\mu(k)
\Big\{-[(2v+h)h+\phi_3^2]c_A(p)c_A(k)+(\phi_1^2+\phi_2^2)\tilde{c}_V(p)c_V(k)\Big\}\, ,
\label{Higgscouplings3}
\end{eqnarray}
\noindent 
A few comments are in order.
First, the fact that the gauge-Higgs couplings are modified 
due to the presence of the KK
resonances is not particular of the specific symmetry considered. For
instance, had we 
consider instead SU(2)$_L\times$U(1)$_Y$ we would have also
obtained shifts in the couplings 
which are not the same for the different
components of the Higgs, and in particular we would still have 
$c_V(p)\tilde{c}_V(p)\not = c_A^2(p)$.
Secondly, if we allow for a Higgs bulk profile, $f_H(z)$, 
the couplings  $c_V$, $\tilde{c}_V$  and $c_A$ would 
depend on this profile. However, since the  Higgs must be quite
localized near the IR brane, the approximation made here (perfect IR
localization) should capture the essence of the effects up to small corrections.
Also, the couplings in (\ref{defgammabeta}) entering in the cubic
and quartic interactions of (\ref{Higgscouplings1}) and
(\ref{Higgscouplings3}) introduce an additional dependence on the
external momentum. 

Finally, the effective low energy theory is obtained by integrating the 5D bulk, 
i.e. it is taking into account the effects of all the KK modes. It is also interesting to 
obtain the effective couplings $c_V$, $\tilde{c}_V$ and $c_A$ by integrating one or two KK modes and 
see how rapidly the process converges. This can be seen in Table~\ref{kktable}, where we show the effective 
couplings at zero momentum 
for the full 5D bulk integration, the case when only one KK mode is integrated out
and taken into account, and finally the results obtained with the first two KK modes integrated out.  
These results are approximated (for instance, we assume $(M^{(1)}_{KK})^2\simeq 6/z_1^2$, and 
$(M^{(2)}_{KK})^2\simeq 30/z_1^2$), 
but already give a sense of the convergence of the 
procedure. 
\begin{table}[h]
\begin{center}
  \begin{tabular}{| c | c | c | c |} 
\hline
   Eff. Couplings & Holography & 1st KK &1st+ 2nd KKs\\ 
\hline
\hline
   $c_V(0)$ & $\frac{2\tilde{g}_5^2 kv^2z_1^2+8}{(\tilde{g}_5^2+g_5^2)kv^2z_1^2+8}$
	       & $\frac{2\tilde{g}_5^2 kv^2z_1^2+12}{(\tilde{g}_5^2+g_5^2)kv^2z_1^2+12}$ 
    	       & $\frac{2\tilde{g}_5^2 kv^2z_1^2+10}{(\tilde{g}_5^2+g_5^2)kv^2z_1^2+10}$\\ 
\hline
   $c_A(0)$ & $\frac{8}{(\tilde{g}_5^2+g_5^2)kv^2z_1^2+8}$ 
	       & $\frac{12}{(\tilde{g}_5^2+g_5^2)kv^2z_1^2+12}$ 
	       & $\frac{10}{(\tilde{g}_5^2+g_5^2)kv^2z_1^2+10}$\\ 
\bottomrule
  \end{tabular}
\end{center}
\caption{Effective couplings computed integrating out the 5D bulk, only the first KK resonance, and 
only the first and 
second KK resonances. The remaining coupling is given by $\tilde{c}_V^2 = 2-c_V^2$~.}
\label{kktable}
\end{table}
The effect of the KK modes comes essentially from the mixing of the axial-vector combination with the 
zero-mode gauge bosons, triggered by the Higgs VEV. We conclude that the KK picture is not a 
bad approximation and the first KK modes do capture the correct
physics in approximate magnitude and sign. 
However, and since it is fairly straightforward to obtain 
the full 5D bulk integration of the holographic picture, we will use
the full result obtained in Sections~\ref{leet} and~\ref{ghint}.

\section{Higgs Contributions to Electroweak Parameters}
\label{higgsconts}

Since the couplings between the Higgs sector and the SM gauge bosons are
modified by the presence of the 5D bulk, we expect effects 
in the electroweak parameters with respect to the SM. In the SM, the
Higgs contributions
to the $S$ and $T$ parameters are finite because the potentially
divergent terms cancel when we add the different diagrams. As we will show,
in the present model the Higgs contribution to $S$ is cutoff sensitive.
In the effective theory described in the previous section, 
the shifts in the Higgs
couplings to the SM gauge bosons will result in additional
contributions to $T$ and $S$ and in particular, in divergent
contributions to $S$.  
Due to the custodial symmetry, there is no tree-level contribution 
to $T$, as can be seen from eq.~(\ref{l2effLB}), since 
$\Pi_{11}=\Pi_{33}=\Pi_L$ at this order. 
In the Appendix we explicitly show that the one-loop 
contribution to $T$ is finite, as expected also from the custodial
symmetry, as well as from  the absence of a counter-term. 
In what follows we present the calculation of the one loop Higgs
contributions to $S$ in the low energy effective theory. 

\subsection{Contribution to $S$ in the Effective Theory}
\begin{figure}
\begin{center}
\epsfig{file=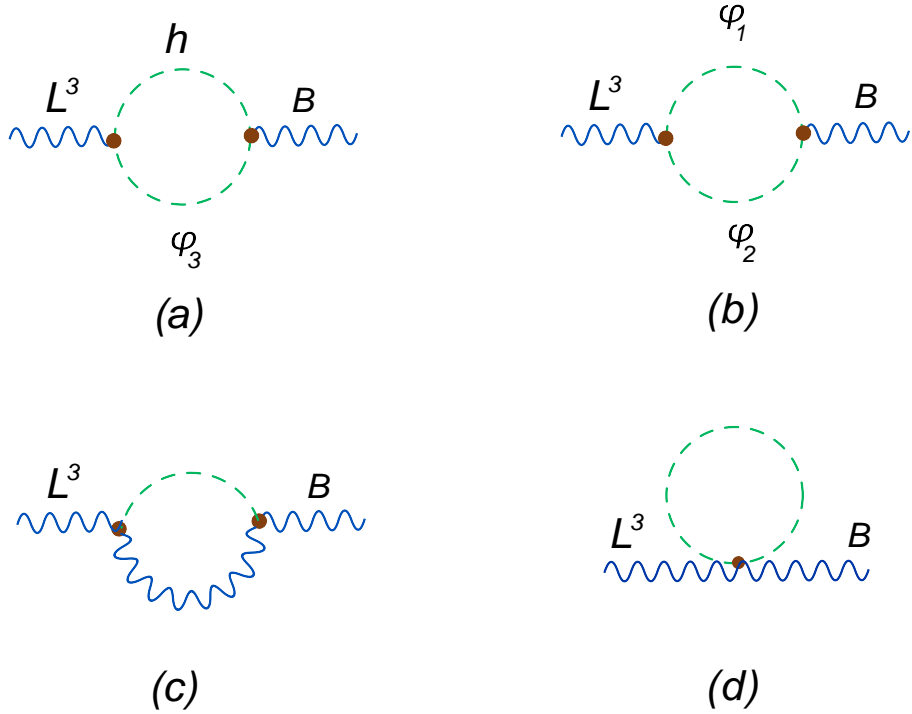,width=12cm}
\caption{One-loop Feynman diagrams contributing to the $S$ parameter
involving the Higgs sector. The large dots stand for the
effective couplings of eqs.~(2.30) and~(2.32).}
\label{figHiggs}
\end{center}
\end{figure}

The one loop contributions of the Higgs sector  to the $S$
parameter are those depicted in the Feynman diagrams of
Figure~\ref{figHiggs}. 
The dots denote the effective couplings in eqns.~(\ref{Higgscouplings1}) and
(\ref{Higgscouplings3}), obtained by integrating out the 5D bulk. 

Although the contributions to one-loop self-energies coming
from the gauge sector are generically gauge dependent, 
the contributions from the diagrams in Figure~\ref{figHiggs} 
{\em to oblique electroweak corrections}  are separately
gauge-invariant. In general, the gauge dependence of gauge-boson
self-energies is cancelled by 
vertex and box diagrams which induce pinch propagator-like 
contributions~\cite{pt}. However, the pinch contributions that 
affect the diagrams of Figure~\ref{figHiggs} are {\em non-oblique}~\cite{papasolo}, 
implying that the oblique pieces of these diagrams are gauge
invariant. Therefore, the contributions of the diagrams of
Figure~\ref{figHiggs} to oblique electroweak parameters are separately
gauge invariant.

Not all contributions from the diagrams in Figure~\ref{figHiggs} should be
considered as contributions to $S$. Some of them are renormalizing 
the Higgs VEV. In order to see how to identify these pieces, it is
instructive to first turn to the tree-level contribution to $S$,
$S_{\rm tree}$, as shown in (\ref{stree}). Since the Higgs is
localized on the IR brane, the effects of EWSB must go through the 
5D bulk in order to be felt by the UV fields. In particular, the
mixing between $B$ and $L_3$ caused by the Higgs VEV in the IR brane, 
picks up  a momentum dependence in the bulk, resulting in {\em
kinetic} mixing, and therefore in a contribution to $S$ given by
$S_{\rm tree}$. The loop contributions in Figure~\ref{figHiggs}
are also IR-localized. External momentum dependence arises from 
either external momentum in the loop, or the momentum dependent
coefficients $c_A(p)$, $c_V(p)$ and $\tilde{c}_V(p)$ appearing in the 
Higgs couplings to gauge bosons in the effective theory.
The latter, is the momentum dependence that the IR-localized loops
acquire when going from the IR brane to the UV, where the elementary 
gauge bosons are. These contributions do not have external momentum
dependence themselves in the IR, and they correspond to various
renormalizations, such as the renormalization of $v$ appearing in
$S_{\rm tree}$ in eqn.~(\ref{stree}). Thus, as a general rule, 
genuine contributions to the $S$ parameter are those with external momentum 
actually flowing through the loop. This amounts to computing the loop
diagrams with the effective couplings $c_A(p)$, $c_V(p)$ and
$\tilde{c}_V(p)$ evaluated at zero external momentum. This means, for instance, that 
the diagram \ref{figHiggs}-(d) will not contribute to $S$, but that
the momentum dependence from the couplings results in a
renormalization of $v$ appearing in $S_{\rm tree}$. 

The diagram \ref{figHiggs}-(c) gives a finite contribution to
$S$. This can be seen by noticing that the corresponding 
loop diagram gives 
\be
i\Pi^{1(c)}_{3B}(p^2) = 
-\frac{\left(g^2+g'^2\right)^2}{4}\,v^2\,c_w\,s_w\,c_A^2(p)\,
\int\frac{d^4k}{(2\pi)^2}\,c_A^2(p-k)\,G_h(k)\,G_Z(p-k)~,
\label{pi1ca}
\ee
corresponding to the $g_{\mu\nu}$ coefficient of the diagrams with
$L^3$ and $B$ inside the loop, and where $c_w$ and $s_w$ stand for the
cosine and sine of the Weinberg angle respectively. 
In the SM limit, $c_A\to 1$, these loop diagrams 
result in finite contributions to $S$ since their derivatives with
respect to the external momentum are finite, even if the vacuum
polarizations themselves are divergent. In the present case, however, the
factor $c_A(q-k)$ regulates the vacuum polarization
itself, since in the large momentum limit $c_A(k)\sim e^{-kz_1}$. 
As a consequence, the contributions of Figure~\ref{figHiggs}-(c) to $S$ are not only finite but
further suppressed. We just denote them as $S^{1-(c)}$ for the
remainder of the paper.

The one loop contributions that do result in  divergences in the $S$ parameter are those
from diagrams \ref{figHiggs}-(a) and \ref{figHiggs}-(b). 
In the SM, these diagrams are responsible for the $m_h$ dependence in $S$ 
and are therefore the main source of bounds on $m_h$ from electroweak precision 
bounds. Using dimensional regularization we obtain:
\bear
S_{\rm loop}^{H} =& & \frac{1}{12\pi}\,(N_\epsilon-1)\,\left[c_V(0)\tilde{c}_V(0)
-c_A^2(0)\right]\nonumber\\
 & &+\frac{1}{2\pi}\,\int_0^1 dx (1-x)x\left[ c_A^2(0)\,\ln\left(\frac{\Delta}{\mu^2}\right) - 
c_V(0)\tilde{c}_V(0)\,\ln\left(\frac{M_W^2}{\mu^2}\right) 
\right]
\nonumber\\
&& + {\rm finite~terms},
\label{shdr} 
\eear
where 
\be
N_\epsilon\equiv \frac{2}{\epsilon} - \gamma+1+\ln 4\pi~,
\label{nedef}
\ee
\be
\Delta\equiv xm_h^2 + (1-x)M_Z^2~, 
\label{deltadef}
\ee
$\epsilon=4-d$, $\mu$ is a renormalization scale, 
and the finite terms come from diagram \ref{figHiggs}-(c).
The first term in (\ref{shdr}) is divergent in the low energy effective theory that results from
integrating out the 5D bulk.  The second term\footnote{In this gauge the NGBs 
are massless at this order, and the $M_Z$ and 
$M_W$ dependence in (\ref{shdr}) comes from other loops 
which are finite, such as the one in Figure~\ref{figHiggs}-(c).} gives the $m_h$ dependence to 
the $S$ parameter.
The SM limit corresponds to taking $(c_V, \tilde{c}_V, c_A)\to 1$. From (\ref{shdr}) we see that 
in this limit the divergent term cancels, and the second term results in the Higgs contribution to $S$ 
in the SM: 
\be
S_{\rm SM}^{H} = (1/12\pi)\ln\left(m_h^2/M_W^2\right)+\cdots~.
\ee
Thus, the result of 
taking into account the effects of the 5D bulk (or of the strongly coupled sector) is twofold: 
it makes the $S$ parameter UV-sensitive and it modifies its $m_h$ dependence.

If we regularize the momentum integrals using a cutoff procedure, 
the divergence in (\ref{shdr}) is logarithmic. In this case the contributions to $S$ 
from the Higgs sector can be written as 
\be
S_{\rm loop}^H = \frac{1}{12\pi}\,\left[c_V(0)\tilde{c}_V(0)-c_A^2(0)\right]\ln\frac{\Lambda^2}{\mu^2}
+ {\rm finite ~terms}~, 
\label{shlog}
\ee  
where $\Lambda$ is the cutoff of the low energy effective theory. This should be the local IR 
cutoff, which is warped down to the TeV scale $1/z_1$ from the Planck scale $k$. So we have  
\be
\Lambda \sim \frac{1}{z_1}~.
\label{cutoff}
\ee
We conclude that the $S$ parameter in AdS$_5$ bulk theories is
logarithmically divergent and therefore cutoff dependent.
In order to remove this divergence, the $S$ parameter must be renormalized by choosing a suitable 
renormalization condition.

\subsection{Renormalization of the $S$ Parameter}
The one loop calculation of the contributions to $S$ from the Higgs sector fixes the divergent 
part of the counter-term in the renormalization procedure. However, it does not fix 
the finite parts, for which we need a renormalization condition.
In order to illuminate the discussion we write the $S$ parameter as
\be
S = S_{\rm tree} + \delta S + S_{\rm loop}~,
\label{swct}
\ee
where $\delta S$ is a counter-term. In general, it can be written as
\be
\delta S =  \delta S^{\rm div.} + \delta S^{\rm finite}
\ee
where $\delta S^{\rm div.}$ cancels the divergence in (\ref{shdr}),
and the finite part $\delta S^{\rm finite}$ is only determined by the
renormalization conditions. The resulting renormalized $S$ parameter 
acquires a scale dependence and can be written as 
\be
S(\mu) = S(\mu_0) +
\frac{1}{12\pi}\,\left[c_V(0)\tilde{c}_V(0)
-c_A^2(0)\right]\,\ln\left(\frac{\mu^2}{\mu_0^2}\right)~,
\label{smu}
\ee
where $\mu_0$ is a reference scale and $S(\mu_0)$ is 
\bear
S(\mu_0) &=& S_{\rm tree} - \frac{1}{12\pi}\,\left[c_V(0)\tilde{c}_V(0)
-c_A^2(0)\right] + S^{1(c)}+\delta S^{\rm finite} \nonumber\\
&&~ ~ \nonumber\\
&&+ \frac{1}{2\pi}\int_0^1dx
x(1-x)\left[c_A^2(0)\ln\left(\frac{\Delta}{\mu_0^2}\right) 
-\tilde{c}_V(0)c_V(0)\ln\left(\frac{M_W^2}{\mu_0^2}\right)\right]~,
\label{smu0}
\eear
where $\Delta$ is defined in (\ref{deltadef}). As mentioned
earlier, the last term in (\ref{smu0}) contains the Higgs mass
dependence of the $S$ parameter. 

In order to determine $S(\mu_0)$ we must choose a renormalization
condition, which basically fixes $\delta S^{\rm finite}$. 
In principle, this could be arbitrarily chosen, for instance to match 
the experimentally measured value of $S$ at some energy scale, such as 
\be
S(\mu_0) = S^{\rm exp.}(\mu_0=M_Z)~.
\label{haha}
\ee
However, given that $S^{\rm exp.}(\mu_0=M_Z)\lae 0.1$,  
the choice in (\ref{haha}) amounts to assume a rather efficient cancellation of 
$S_{\rm tree}$ against $\delta S^{\rm finite}$ as well as against the
loop contributions. Although this choice does not result in a numerically fine-tuned
cancellation, it would imply that the leading order contribution to $S$
in the large $N$ expansion, 
\be
S_{\rm tree}\simeq \frac{O(1)}{\pi}\,N~,
\ee
is not a good enough
approximation and that the next order in $N$ is equally important. 
In order to see this, we define the number of colors
$N$ in the 4D CFT in term of the 5D gauge couplings by
\be
\frac{1}{N} = \frac{(g_5^2 + \tilde{g}_5^2)}{16\pi^2}\,k~,
\label{nscaling}
\ee
which reflects that the large $N$ corresponds to the perturbative
expansion in the 5D gauge theory. 
Thus, the loop diagrams considered here are suppressed contributions 
in the large $N$ expansion. This would mean that a renormalization
condition such as (\ref{haha}) implies that an $O(N)$ contribution
such as $S_{\rm tree}$ is efficiently canceled by $O(1)$
contributions coming from loops, which may call into question our use
of the large $N$ expansion, i.e. our use of perturbation theory in the
5D theory. 

As concretes examples, we can study two different limits. First we 
consider $m_1z_1\gg 1$, which corresponds to a heavy Higgs or nearly 
Higgsless scenario.
Using (\ref{stree}), results in~\cite{higgsless} 
\be
S_{\rm tree} \sim \frac{3}{4}\,\frac{N}{\pi}~.
\label{svsn}
\ee
To obtain the one-loop contribution in this limit we need the zero-momentum limit 
of $c_V,\tilde{c}_V$ and $c_A$, which results in 
\bear
\tilde{c}_V(0)c_V(0)& = & \frac{4\left[(1+m_1z_1)\tilde{g}_5^2 +
g_5^2\right]\,\left[(1+m_1z_1)g_5^2 +\tilde{g}_5^2\right]}
{(2+m_1z_1)^2\,(g_5^2+\tilde{g}_5^2)^2}\label{cvcv}\\
c_A^2(0) & = & \frac{4}{(2+m_1z_1)^2}
\label{ca2}~,
\eear
where $m_1$ is defined in (\ref{m1def}). Then, taking $m_1z_1\gg 1$ and using (\ref{nscaling})
we can see that the loop contributions in (\ref{smu}) are of order
$O(1)$ in the large $N$ expansion. We obtain
\be
S_{\rm loop}\sim\frac{1}{12\pi}\,\ln\frac{\mu^2}{m_h^2} \, .
\ee
We can also consider the limit $m_1z_1\ll1$, in which case we have
\begin{eqnarray}
S_{\rm tree} &\sim& 2\pi v^2 z_1^2~,\\
S_{\rm loop} &\sim& \frac{1}{N}\frac{\pi}{3} v^2 z_1^2\,\ln\frac{\mu^2}{m_h^2}\, .
\end{eqnarray}

A more conservative renormalization condition would be 
\be
S(\mu_0) \sim S_{\rm tree}~,
\label{bua}
\ee
which amounts to assume that there is no significant cancellation of 
the tree-level contribution from the counter-term or from loop
corrections. With this choice, a large positive 
$S$ parameter is still predicted, but the prediction cannot be made
precise. 

Although the renormalization condition in (\ref{bua}) avoids  large
cancellations of $O(N)$ and $O(1)$ contributions, the finite pieces of
the counter-term could still affect significantly the loop
contributions. In particular, the last term in (\ref{smu0}) 
containing  the information on the Higgs mass, can be affected by the 
renormalization condition since it is of $O(1)$, just as 
the counter-term is expected to be. We then conclude than in these
theories the $S$ parameter cannot be used to put a bound on the Higgs
mass. 

The ``correct'' renormalization condition might be somewhere in between
these two extremes, i.e. there may be some cancellation of $S_{\rm tree}$
dictated by the unknown UV (or CFT) physics. In any case, what is
clear from our calculation is that the composite, strongly-coupled 
Higgs sector suffers shifts in its couplings to the SM gauge fields in
such a way that the usual cancellations in the diagrams of
Figure~\ref{figHiggs}-(a) and Figure~\ref{figHiggs}-(b) do not occur. 
Thus, this misalignment of the gauge-Higgs couplings with respect to
their SM values, results in a dependence on the cutoff scale (in the 5D
language), or the matching scale with the 4D CFT (in the 4D picture).

\section{Discussion and Conclusions}
\label{conclusions}

We have computed the one-loop contributions to the $S$ parameter 
from the Higgs sector in bulk AdS$_5$ theories of EWSB. In these
generic 5D setups we have used  the minimal extension of the
gauge group that protects isospin symmetry in the bulk, avoiding a
tree-level $T$ parameter. Our results show that the $S$ parameter 
is UV-sensitive and therefore it  must be renormalized. The appearance of
divergences in $S$ are a consequence of the misalignment between the 
gauge fields in the IR, where they interact with the IR-localized
Higgs, and the UV fields which constitute the elementary degrees of 
freedom in terms of the holographic picture. 
This misalignment is produced by the 5D bulk between the IR
and the UV branes, or the strong dynamics  from the 4D CFT, 
and it occurs independently of the choice of bulk
gauge symmetry. These divergences are then completely generic in bulk AdS$_5$ models 
of electroweak symmetry breaking. Their origin is fundamentally 
different from the logarithmic divergence found in Reference~\cite{barbieri}, 
which has origin in the mixing of the Higgs with a state resulting from the symmetry breaking 
pattern in that model. On the other hand, they are similar in spirit 
to the matching-scale dependence found in References~\cite{tsmoose}
and \cite{sally} in a three-site Higgsless
model. 

It is also possible to understand the occurrence of these divergences 
in a generic operator analysis. For instance, the operator
\be
{\cal O}_H = (H^\dagger H)\left|D_\mu H\right|^2~
\label{oh}
\ee
contributes to $S$ when inserted in one-loop diagrams. Its contribution
is logarithmically divergent and results in 
\be
S_{{\cal O}_H} \sim
-\frac{c_H\,v^2}{12\pi}\,\ln\left(\frac{\Lambda^2}{m_h^2}\right)~,
\label{oh2sdiv}
\ee
where $c_H$ is the corresponding coefficient of ${\cal O}_H$. On the
other hand, we can do the matching of this operator to the 
AdS$_5$ bulk theory. Expanding $\Pi_L$ in eqn.~(\ref{l2effLB})  
to fourth order in $v$ at zero momentum, we obtain that
\be
c_H = -\frac{g_5^2 + \tilde{g}_5^2}{4}\,k\,z_1^2~,
\label{chmatch}
\ee
which results in a prediction consistent with (\ref{shlog}). Thus, we
see that the divergence in $S$ is a generic feature in strongly
coupled theories, rather than specific to AdS$_5$ bulk  models. 

Coming back to the AdS$_5$ bulk models discussed in the paper, the 
renormalized $S$ parameter has a calculable scale dependence given
in (\ref{smu}). We discussed the possible choices of renormalization
conditions. Although in principle it is possible to choose an
arbitrary condition, so as to adjust the renormalized value of $S$ to
any desired value, we showed that asking for a significant
cancellation of the tree-level value $S_{\rm tree}$, which is of
$O(N)$ in the large $N$ expansion, might be unnatural if the expansion
is to be trusted. However, and by the same argument, the Higgs mass
dependence in $S$, which appears in (\ref{smu0}), is of $O(1)$ and
therefore can be naturally affected by a 
renormalization procedure triggered by $O(1)$ one-loop corrections. 
We then conclude than in these theories there is no bound on the Higgs
mass that can be extracted from $S$.

We only computed the one-loop contributions from the Higgs
sector. However, we also expect divergent contributions from fermions and gauge bosons.
As we have shown, the divergences can be associated with the shifts in
the couplings between the SM gauge fields localized in the UV and the
composite fields localized towards the IR. Thus, we expect a similar 
effect for composite fermions.
These are more model-dependent and we leave their study for future
work. 

We finally comment on the case where the light fermions are de-localized.
The limit of exact de-localization, results in flat zero modes.  
In this limit~\cite{adms,delocal} there is no tree
level $S$ and we do not expect divergences in the loop contributions.
In this case, since the SM fermions exactly correspond to the zero modes, the
SM gauge fields are also de-localized and their couplings with the Higgs and
fermions are canonical.

\section*{A.~Higgs Sector Contribution to $T$}
\label{appendixT}
\begin{figure}
\begin{center}
\epsfig{file=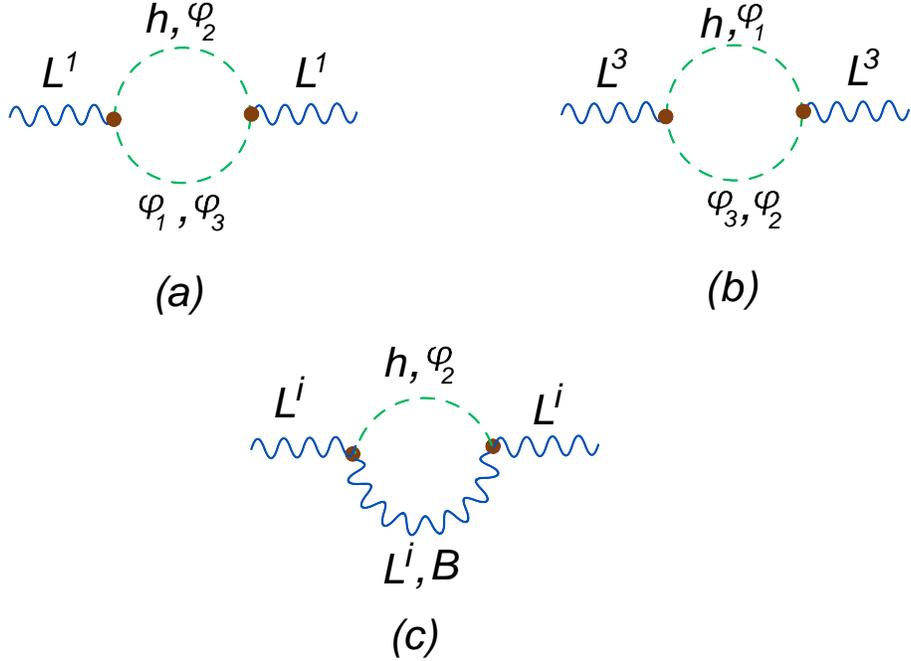,width=12cm}
\caption{Relevant one-loop diagrams contributing to the $T$ parameter
involving the Higgs sector. Here $L^{i}=L^1, L^3$.}
\label{higgs2t}
\end{center}
\end{figure}

\renewcommand{\theequation}{A.\arabic{equation}}
\setcounter{equation}{0}
Here we compute the one-loop contributions to the $T$ parameter
coming from the Higgs sector.  
The relevant diagrams 
are shown in Figure~\ref{higgs2t}. The contributions to 
$T\propto\Pi_{11}(0)-\Pi_{33}(0)$ are similar to the case of the SM,
but changing the usual interactions by those of
eqs.~(\ref{Higgscouplings1}-\ref{Higgscouplings3}). The effective couplings 
$c_V(p)$, $\tilde{c}_V(p)$ and $c_A(p)$ associated to the external legs 
are evaluated at zero momentum. 
Notice that the NGBs $\phi_i$ are degenerate at tree level. From 
eq.~(\ref{Higgscouplings1}) we can see that the one-loop contributions 
to $T$ from the Feynman diagrams~\ref{higgs2t}-(a) and~\ref{higgs2t}-(b) 
exactly cancel. 
We compute now the diagram in Figure~\ref{figHiggs}-(c).
In order to see explicitly that this contribution is finite, we work in the
diagonal basis $\{\gamma,Z\}$. There are two diagrams contributing to
$\Pi_{11}$, one with a Higgs field $h$ and another with a NGB 
field $\phi_2$ propagating in the loop. This gives:
\begin{eqnarray}
i\Pi_{11}(0) &=&\frac{g^4\,v^2}{16}\,c^2_A(0)\,
\int \frac{d^4k}{(2\pi)^4}\,c^2_A(k)\,G_hG_L \nonumber\\
+\frac{g^2g'^2v^2}{16}\,
\int\frac{d^4k}{(2\pi)^4}&& 
\frac{\left[g_5^2c_A(0)\left(c_V(k)+\tilde{c}_V(k)\right)+\tilde{g}_5^2c_A(k))\right]^2}
{\left(g_5^2+\tilde{g}_5^2\right)^2}\label{pi11}\\
&&\times G_{\phi_2}\,(c_w^2G_\gamma+s_w^2G_Z)
\, ,\nonumber
\end{eqnarray}
\noindent where $G_{h,\phi_i}$ are the propagators of the
Higgs and the NGBs and
we have factorized the gauge propagators as 
$G^A_{\mu\nu}=P_{\mu\nu}G_A$. 
On the other hand, the contribution to $\Pi_{33}$ is given by:
\begin{eqnarray}
i\Pi_{33}(0)&=&\frac{g^4\,v^2}{16}\,c^2_A(0)\,
\int \frac{d^4k}{(2\pi)^4}\,c^2_A(k)\,G_hG_L \nonumber\\
&+&\frac{g^2g'^2\,v^2}{4}\,c_A^2(0)\,
\int\frac{d^4k}{(2\pi)^4}\,c^2_A(k)\,G_h\,\left[ s_w^2 G_Z + c_w^2 G_\gamma\right]
\,.
\label{pi33}
\end{eqnarray}
We can see that in the one-loop contributions to 
the $T$ parameter, proportional to $\Pi_{11}(0)-\Pi_{33}(0)$, the first terms in 
(\ref{pi11}) and (\ref{pi33}) cancel. Also, and just as in the case 
of the discussion of $S^{1-(c)}$ in Section~3.1, the vacuum polarizations are finite.
This is because 
$f_{V,A}(k,z_1)$ are exponentially suppressed at large momentum, $\sim
e^{-kz_1}$, implying that also $c_V(k)$, $\tilde{c}_V(k)$ and $c_A(k)$  are. 
The exponential suppression is due
to the Higgs localization in the IR boundary. Had we considered a
Higgs with a profile in the bulk, we would have obtained a power suppression. 
All the gauge propagators can be approximated at large momentum,
$kz_1\gg 1$, by $1/(k^2\log k)$. 
Therefore the integrands of eqns.~(\ref{pi11}) and (\ref{pi33}) are exponentially
suppressed for large momentum and the contribution to $T$ from the Feynman 
diagram~\ref{figHiggs}-(c) is finite. 

There are also contributions with only gauge fields running in the
loops. Since the difference between the vector and
axial correlators is exponentially suppressed at large momentum, these
contributions to $T$ are also finite.

\bigskip

\noindent
{\bf Acknowledgments:}
We thank Sekhar Chivukula, Roberto Contino, Elizabeth Jenkins, Ryuichiro Kitano, 
Aneesh Manohar, \'Alex Pomarol and Erich Poppitz for stimulating discussions. 
We also acknowledge the support of the State of S\~{a}o Paulo
Research Foundation (FAPESP). G.B. also thanks the Brazilian  National Counsel
for Technological and Scientific Development (CNPq) for partial support.



\end{document}